# AES-CBC Software Execution Optimization

Razvi Doomun*, Jayramsingh Doma, Sundeep Tengur
*Computer Science and Engineering,*
*University of Mauritius*
r.doomun@uom.ac.mu, kartouss@gmail.com, tempo14@gmail.com

**Abstract**

*With the proliferation of high-speed wireless networking, the necessity for efficient, robust and secure encryption modes is ever increasing. But, cryptography is primarily a computationally intensive process. This paper investigates the performance and efficiency of IEEE 802.11i approved Advanced Encryption Standard (AES)-Rijndael ciphering/deciphering software in Cipher Block Chaining (CBC) mode. Simulations are used to analyse the speed, resource consumption and robustness of AES-CBC to investigate its viability for image encryption usage on common low power devices. The detailed results presented in this paper provide a basis for performance estimation of AES cryptosystems implemented on wireless devices. The use of optimized AES-CBC software implementation gives a superior encryption speed performance by 12 – 30%, but at the cost of twice more memory for code size.*

## 1. Introduction

Wireless technologies have exploded into prominence over the last few years, with newer and more advanced standards emerging all the time. As wireless traffic becomes more pervasive, the requirement for high quality security becomes even more important. All the recent standards have included security from the start, with the older standards like IEEE 802.11 being brought up to date with its 802.11i security extensions. Users are expecting to secure data transmission and storage on wireless mobile devices, which require efficient cryptographic algorithms. Hence, there is increasing need for efficient AES-CBC software implementation since it has become a key ingredient in IEEE 802.11i wireless security protocol [1]. Many of the IEEE standards, such as 802.11, 802.15 and 802.16 use AES-CCM [2][1] as the basis for their security; and this is a good choice because it provides both encryption and authentication in a single efficient solution. Security and performance considerations are therefore both imperative during the design phase of encryption algorithms.

With the advent of AES in IEEE 802.11i, as considered in this work, and the high prospects of wireless systems, this research holds a prominent position in the evaluation and analysis of the structure of the Rijndael algorithm (AES-CBC) from the resource constraint wireless systems point of view. Many cryptographic algorithms, such as AES-Rijndael, which are compact and efficient to implement on high-performance microprocessors, may not be implementable efficiently on smaller and less powerful microprocessors found in low-power mobile devices. A resource-constrained wireless system's efficiency is invariably related to the size of the code [3]. The efficiency of a program increases, as the code size decreases and the execution speed increases. Therefore, implementation of cryptosystems on with very tight memory constraints mobile devices introduces new challenges. In [4], the authors investigate the speed measurement of several cryptographic system libraries to determine if they are feasible for Palm devices or if they are too complex. The main finding is that it can be valuable to encrypt real-time data if pre-computation of certain steps is allowed but it may cause a problem for devices with very limited space. Hence, optimization is often possible through a closer inspection of the encryption software algorithms.

The security suites can be more broadly classified by their properties: encryption only (AES-CTR), followed by authentication only (AES-CBC MAC), and finally encryption and authentication (AES-CCM) [1]. AES-CTR (counter mode of cryptographic operation with AES) means that the CTR mode uses AES as the block cipher; and provides access control, data encryption and optional sequential freshness. Authentication is done using the cipher block chaining with message authentication code (CBC-

MAC), which creates a message integrity code using a block cipher in CBC mode, and computes a MAC over the packet and includes the length of the authenticated data. The code can be computed upon packet reception and can be compared with the one received.

AES-CCM is a combination of the encryption and authentication suites detailed above. It has three inputs; the data payload to be encrypted and authenticated, the associated data (header etc.) to be authenticated only, and the nonce to be assigned to the payload and the associated data [1]. There are varying MAC lengths to choose from for AES-CBC-MAC and AES-CCM modes of operation (4, 8 or 16 bytes), allowing for some scalability of security depending on application requirements.

In this paper, we study the performance of AES-CBC software execution and its operation complexity. An overview of Rijndael algorithm and complexity is presented in section 2. The CBC mode is described in section 3. Simulation results are analyzed and discussions are provided in section 4 and 5 respectively. Finally, we conclude and summarize the main findings of the paper.

## 2. Rijndael block cipher algorithm

The Advanced Encryption Standard (AES)- Rijndael [5][6] is an iterated symmetric key block cipher with a variable block length and a variable key length that can be independently specified to 128, 192 and 256 bits. The four main functions that comprise the AES algorithm are Add Round Key, Substitute Byte, Shift Rows and Mix Columns. A data block to be processed by Rijndael is partitioned into an array of bytes, called State, and each cipher operation is byte-oriented. Add Round Key, the first step in the transformation, performs a bitwise XOR of the state and the round key matrix. This transformation is its own inverse. Then, the Sub-Byte transformation is a non-linear byte substitution operation that is composed of two sub-transformations: multiplicative inverse and affine transformation. In typical software implementations [7], these two sub-steps are combined into a single table lookup called substitution box or S-box. The ShiftRows function is a linear diffusion process that operates individually on the last three rows of the state matrix. A simple byte transposition cyclically shifts to the left the bytes in the rows by an offset varying from one to three. The second, third and four rows' elements are shifted one byte, two bytes and three bytes to the left, respectively. For full encryption, the data is passed through Nr rounds (Nr=10, 12, 14) that are governed by the four transformations, as shown in figure 1.

Most software based AES implementations are written in the C/C++, assembler or Java programming languages. Since these implementations are based on different APIs, processors, compilers and various design assumptions, they are hard to compare. Some researchers however, have tried to compare the AES performance in an efficient way on various platforms.

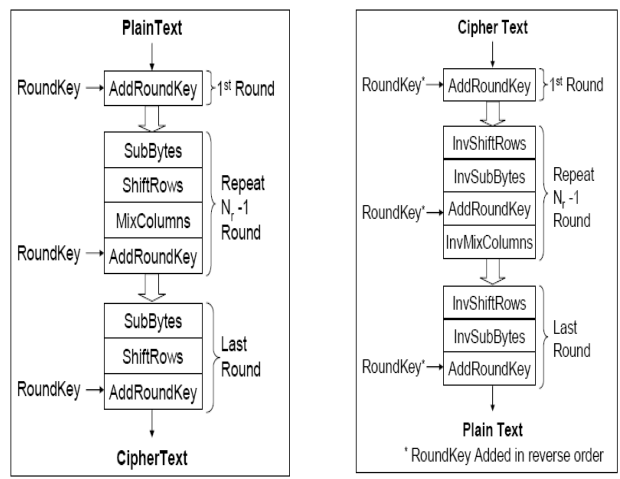

**Figure 1 AES transformation**

Paper [8] claims that the fastest software based AES implementation requires 237 cycles to encrypt one data block. This result was obtained on a Pentium II 450 MHZ platform and the software implementation was optimized on the basis of several large pre-calculated tables. Furthermore, the work also assumed that all data variables are directly available.

**AES-Rijndael computational analysis**

There are two methods commonly used in order to distinguish between time critical operations and non time-critical AES operations [10]. The first method is based on analyzing AES transformations on arithmetical or mathematical level, while the second method is based on analyzing the transformations on the amount of executed instructions. The basis of the first method is: what are the arithmetical operations and how many clock cycles will these operations require. The second method will be performed by using a simulator that will give a detailed profiling information of the executed instructions.

An mathematical analysis of AES computational cost is given in [9][10], and each AddRoundkey is implemented with $8N_b$ bytewise-ANDs and $4N_b$ bytewise-ORs, where is $N_b$ = block length/32, each SubByte operation incurs $3N_b$ bytewise-ANDs and

$2N_b$ bytewise-ORs, each ShiftRows consists of $3N_b$ shifts of bytes and $3N_b$ bytewise-ORs and each round operation with $19N_b$ bytewise-XORs, $8N_b$ bytewise-ORs and $64N_b$ shifts; or $38N_b$ bytewise-ANDs, $27N_b$ bytewise-ORs and $64N_b$ shifts. AES for one block of data is a function of the block size, the key size, the number of encryption rounds ($N_r$) and the number of processing cycles required for performing basic operations bytewise-AND ($T_a$), bytewise-OR ($T_o$), and bytewise shift ($T_s$) and expressed in general terms as:

$T_{\text{AES-ENCRYPT}} = (46N_b\ N_r - 30N_b)T_a + [31N_b\ N_r + 12(N_r - 1) - 20N_b\ ]T_o + [64N_b\ N_r + 96(N_r - 1) - 61 N_b]T_s$.

A significant limitation of AES is related to its decryption because the cipher and its inverse make use of partially different code. The decryption code has Inverse MixColumns operation, which uses a transformation with another polynomial, $0Bx^3 + 0Dx^2 + 09x + 0E$. This leads to extra processing complexity for decryption as multiplication by bigger coefficients is more complex. The difference in computation between one Inverse MixColumn and MixColumn operation is $[96N_bT_a + 72N_bT_o - 32N_bT_s]$.

Therefore, the total number of processing cycles in computational effort required for AES decryption of **one block of data** is given:

$T_{\text{AES-DECRYPT}} = T_{\text{AES-ENCRYPT}} + \{[96\ N_bT_a + 72\ N_bT_o - 32\ N_bT_s] \times (N_r - 1)\}$.

On analyzing AES transformations on arithmetical level, showed that the MixColumn and the Inverse MixColumn transformation are the most time-critical operation. The second method, which was based on the simulation tests, showed that the MixColumn and the Inverse MixColumn transformation contain the most executed instructions, and that most of these instructions are related to integer computations. Therefore one can conclude that the MixColumn and Inverse MixColumn transformation are indeed the most time-critical operations.

## 3. CBC mode

The Cipher Block Chaining (CBC) [12] is a popular block cipher mode operation where each plaintext is XORed with the previous ciphertext block before being encrypted. Hence, each ciphertext is dependent on all plaintext blocks up to that stage. A plaintext message $M$ is divided into $t$ $n$-bit blocks $M_i$ and the ciphertext $C_i$ is given as:

$C_i = E_k(M_i \oplus C_{i-1})$, $i = 1, 2, ...t$.

In the CBC mode, the value $C_{i-1}$ is used to randomize the plaintext by combining with data blocks $M_i$ to hide patterns and repetitions. To enable the encryption of the first plaintext block ($i=1$), $C_0$ is defined as the Initial Value (IV), which should be randomly chosen and transmitted securely to the recipient.

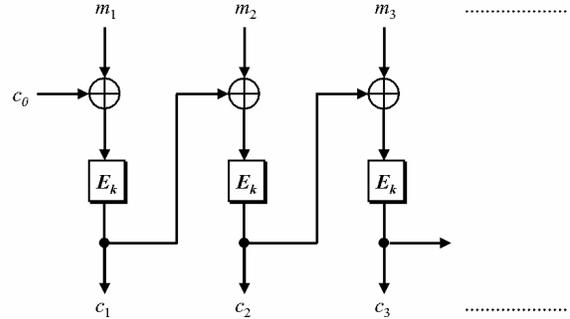

**Figure 2: CBC model**

CBC mode is as secure as the underlying block cipher against standard attacks. In addition, any patterns in the plaintext are concealed by the XORing of the previous ciphertext block with the plaintext block. Note also that the plaintext cannot be directly manipulated except by removal of blocks from the beginning or the end of the ciphertext. The initialization vector should be different for any two messages encrypted with the same key and is preferably randomly chosen. It does not have to be encrypted and it can be transmitted with (or considered as the first part of) the ciphertext. CBC overcomes the security deficiency of the electronic codebook mode; the input to the encryption algorithm consists of the XOR of the plaintext block and the ciphertext produced from the previous plaintext block as illustrated in figure 2. This makes it more difficult for a cryptanalyst to beak the code using strategies that look for patterns in the ciphertext, patterns that may correspond to the known structure of the plaintext. With this chaining scheme, the ciphertext block for any given plaintext block becomes a function of all the previous ciphertext blocks.

## 4. Simulation results

Software encryption is still being widely used due to the software features of portability and flexibility. However, unoptimized software encryption is very slow and is insecure in many aspects of key management and program manipulation. The coding of the program was performed in the C language, which is a high-level language defined at higher abstract levels and is programmer-friendly. The high-level language needs to be compiled into a low-level language before execution. The main advantage of using a high-level language is code portability i.e. the ability of the code to be transferred to a different system or environment with minimal amounts of modification and redevelopment.

Modularization of encryption software is the technique of splitting a large program into smaller modules. The advantage of modularization is the ease of maintenance and code debugging. Modularization helps in code-reuse, which reduces run-time memory. A cryptographic system is developed as a separate module with sub-modules implementing the details. In the C language, modularization is achieved by dividing the code into various functions. When the embedded system needs to encrypt or decrypt data it invokes the corresponding module, which executes its tasks and then returns the output to the host function. The main program was divided into different modules termed encrypt(), decrypt()and KeyExpansion() functions.

**Optimization**
Maintaining pre-computed tables to simplify program operations and improve performance is a common practice. For AES, a method to combine different operations of the round transformation in a single set of table lookups was suggested in [11]. This approach basically combines the matrix multiplication required in the *MixColumn* operation with the *S-box*, and involves 4 tables with 256 4-byte entries, i.e., 4 KByte of memory. Because encryption and decryption must use different tables, the total memory requirement amounts to 8 KByte. Another solution is to trade memory for speed, and use two 256-byte lookup tables for the *SubByte* and *InvSubByte* operations, while implementing the *Mix-Column/Inverse MixColumn* operations separately. Here, again, various trade-offs are possible. Each call to the *MixColumn* or *InvMixColumn* operations results in sixteen field multiplications. A straightforward implementation of the multiplication operation in the field is MIPS-intensive. Since one of the multiplicands is fixed (with values limited to 6 field elements, i.e., *f*01*g*; *f*02*g* and *f*03*g* for *MixColumn* and *f*0b*g*; *f*0d*g*, and *f*09*g* for *InvMixColumn*), a conventional field multiplication operation can be replaced by a table lookup, requiring a new 6x256 table, each element of which is 8-bit wide. Image specifications used for simulation are summarized in table 1.

**Table 1: Image Type and Size**

| Image | Details | | |
|---|---|---|---|
| | Type | Resolution | FileSize |
| Image 1 | 24-bit Bitmap | 200 x 200 | 117 KB |
| Image 2 | 24-bit Bitmap | 300 x 300 | 263 KB |
| Image 3 | 24-bit Bitmap | 400 x 400 | 468 KB |

Recursive tasks have an overhead that needs to be checked when the instruction sequence should jump out of the loop. For a small number of repetitions, the overhead could be removed altogether by replacing the loop with the code components for that fixed number of times. This technique is called loop unrolling. When two loops are being executed with similar tasks that can be sequentially adjusted, it is better to combine the two loops into a single loop. This technique is called loop merging. This reduces the total overhead time of executing multiple loops to the overhead of a single loop.

AES-ECB with increasing key size/rounds e.g. 128/10, 192/12 and 256/14 causes: AddRound Key execution time to increase by 13 – 16 %; SubBytes execution time increases by 17 – 22%; ShiftRow execution time increases 20 – 22 % and Mixcolumn execution time increase 21 – 25%. When the number of AES rounds is increased by 2 stepwise (i.e. 2, 4, 6, 8…), the encryption time for a data block of 16 bytes is augmented by a margin of 14-19% whereas a 15-30% rise is observed in the decryption time.

After software optimisation techniques were applied:
- SubBytes()showed a performance gain of 20 % in terms of execution speed.
- ShiftRows() displayed a performance of 30 % in terms of execution speed.
- AddRoundKey() displayed a performance gain of 21 % in terms of execution speed.
- Mixcol() displayed a performance gain of 13 % in terms of execution speed.

This shows that a 20 – 30 % performance gain was obtained by optimizations for encrypt() function ShiftRows() experienced the best ratio of optimisation than all other sub-functions. MixColumns() was one of the least optimized function. This was due to the implementation of a look-up table for the GF multiplication. . Simulations were performed on a Pentium-4 3.0 GHz with 512 MB DDR RAM. Figure 3 shows the encryption time results for different optimization scenario and cipher parameters. We examined the performance of encryption/decryption by optimizing the round transformation operations. Alternate round execution gives a trade-off between minimizing code size and reducing execution time.

AES-ECB 128/10 rounds partially optimized with alternate rounds (Opt1) gives 13% less encryption time compared to original unoptimized. AES-ECB 128/10 partially optmized with two alternate rounds (Opt2) gives a performance gain of 12 % on encryption time. AES-ECB 128/10 fully optimized (OptF) improved encryption time by 20% but with memory storage for code size that is doubled.

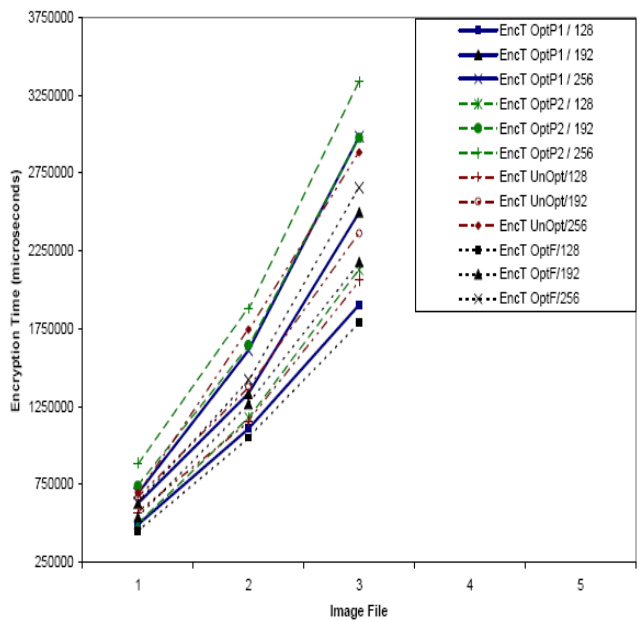

**Figure 3: Encryption time v/s varying image size (117, 263, 468 Kbytes) for different key size 128/10, 192/12, 256/14 and optimization scenarios.**

## 5. Results discussion

In many cases, programs have a high-cost critical path that needs to be optimized. It is therefore necessary to optimize the critical paths to a higher extent than the less critical paths. A complex mathematical can be made simple by dividing it into smaller components so that they can be executed in parallel. Observations from the various stages of the code optimizations revealed that the **MixColumns()** function was consuming more time than other sub-modules combined in the **encrypt()** function. This was due to the **mul()** function in the **MixColumns()** function, which was used to perform the Galois Field (**GF**) multiplication on the data operands. GF multiplication was performed by implementing a look-up table to defeat any timing attacks. The results reveal that **decrypt()** takes more time than **encrypt().** This is due to the added complexity of the GF multiplication in **InvMixColumns()** of **decrypt().** The **InvMixColumns()** needs to perform four multiplications while the **MixColumns()** needs to perform only two multiplications per each byte of the **state**. Complex functions in AES can be made simpler by exploring other alternatives such as look-up tables and bit-manipulation. The **SubBytes()** can be implemented by using  the formula but it consumes lot of processor cycles. So, **SubBytes()** was implemented by using a look-up table.  Some tasks in a program need to be executed a finite number of times. Recursive tasks have an overhead that needs to be checked when the instruction sequence should jump out of the loop. For a small number of repetitions, the overhead could be removed altogether by replacing the loop with the code components for that fixed number of times. This technique is called loop unrolling. Code Sample A presents the loop unrolling for **AddRoundKey()** and Code Sample B presents loop unrolling for **SubBytes().**

**Initial code:**
int i, j;
for (i=0; i<4; i++)
for(j=0; j<4; j++) a[i][j] ^= rk[i][j];

**Modified code:**
int i;
for (i=0; i<4; i++)
{
a[i][0] ^= rk[i][0];
a[i][1] ^= rk[i][1];
a[i][2] ^= rk[i][2];
a[i][3] ^= rk[i][3];
}
Code Sample A: Loop unrolling for **AddRoundKey( )**

**Initial code:**
int i, j;
for (i=0; i<4; i++)
for(j=0; j< BC; j++)
a[i][j] = box[a[i][j]];

**Modified code:**
int i;
for (i=0; i<4; i++)
{
a[i][0] = box[a[i][0]];
a[i][1] = box[a[i][1]];
a[i][2] = box[a[i][2]];

a[i][3] = box[a[i][3]];
}
Code Sample B: Loop unrolling for **SubBytes( )**

When two loops are being executed with similar tasks that can be sequentially adjusted, it is better to combine the two loops into a single loop. This technique, called loop merging, reduces the total overhead time of executing multiple loops to the overhead of a single loop. Code Sample C presents the loop unrolling and merging with constants substitution for **ShiftRows( ).**

**Initial code:**
```
int i, j;
for (i=0; i<4; i++)
{
for(j=0; j< BC; j++)
tmp[j] = a[i][(j + shifts[BC-4][i]) % BC];
for(j=0; j< BC; j++) a[i][j] = tmp[j];
}
```

**Modified code:**
```
int i;
for (i=1; i<4; i++)
{
tmp[0] = a[i][(0 + i) % BC];
tmp[1] = a[i][(1 + i) % BC];
tmp[2] = a[i][(2 + i) % BC];
tmp[3] = a[i][(3 + i) % BC];
a[i][0] = tmp[0];
a[i][1] = tmp[1];
a[i][2] = tmp[2];
a[i][3] = tmp[3];
}
```
Code Sample C: Loop unrolling and merging

To maintain a balance between optimization and code size, only internal loops have been unrolled. Loop unrolling makes the code run smoothly in low processor-intense devices while monitoring the code size ensures efficient use of memory. This is because small mobile devices such as smart phones and PDAs have small embedded memory and might not be able to fit the program in the memory if the code is bulky. AES encryption and decryption code size for optimized and unoptimized versions are given in table 2.

**Table 2: Unoptimised v/s Optimised AES Code Size**

| AES Encryption Operations | Unoptimised code /bytes | Optimised code / bytes |
|---|---|---|
| KeyExpansion | 1016 | 1703 |
| AddRoundKey | 139 | 443 |
| SubBytes | 142 | 476 |
| ShiftRows | 479 | 583 |
| MixColumns | 277 | 2533 |
| AES Decryption Operations | Unoptimised code /bytes | Optimised code / bytes |
| KeyExpansion | 1016 | 1703 |
| AddRoundKey | 139 | 443 |
| InvSubBytes | 142 | 476 |
| InvShiftRows | 479 | 572 |
| InvMixColumns | 317 | 3374 |

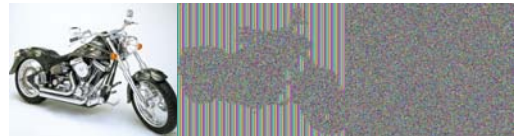
(a)  (b)  (c)
Figure 4: (b) AES-ECB v/s (c)AES CBC

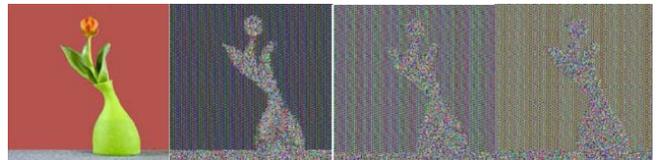
Figure 5a: AES-ECB using increasing number of rounds (2,4,10)

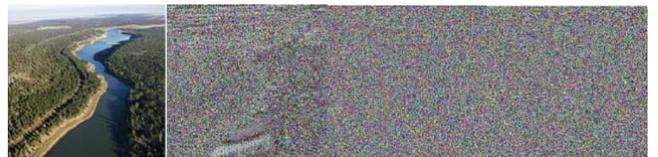
Figure 5b: AES-ECB using varying number of rounds (2, 4, 10)

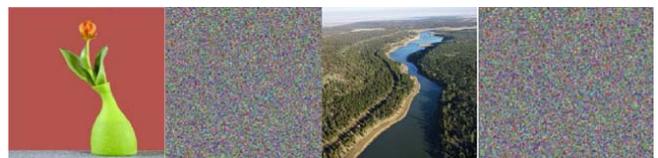
Figure 5c: AES-CBC for any number of rounds

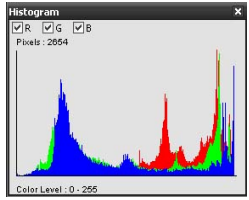

(a) before encryption

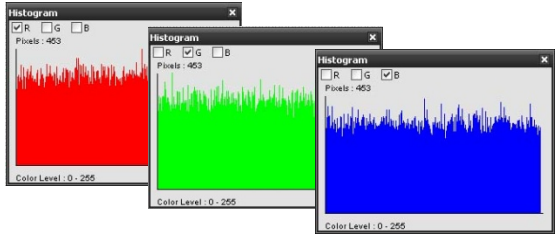

(b) after encryption

**Figure 6: Image histograms before and after encryption**

When the image contains homogeneous texture zones, all the identical blocks having same content also produce same output after AES-ECB ciphering. The AES-ECB mode of operation failed to conceal all the hidden details (visual texture patterns) of a bitmap image and AES-CBC had the highest encryption time. Hence, the AES-ECB encrypted image will also contain textured zones and the entropy of the image is not maximum, as shown in figure 4 and figure 5. As the number of AES-ECB encryption rounds increases, the visual texture patterns are concealed to higher degree. An ideal encrypted image shows a flat histogram distribution of pixels values. The image histogram in figure 6 shows the distribution of RGB pixel values for the unencrypted image and the three encrypted colour channels. Various colour peaks can be noted, showing that textures and visual components exist in the unencrypted image. Previous simulations have demonstrated that the AES algorithm performs well for encryption and the resultant images are scrambled. However, it is true that for some types of images, mainly single objects on a plain monochrome background, the AES-ECB encrypted image still holds some visual information about the original image such as shape and texture in some cases.

## 6. Conclusion

There are many design tradeoffs to consider when implementing the AES-CBC algorithm in software. In resource-constrained platform, the memory requirements, power consumption and throughput are important considerations. The AES-ECB and CBC image encryption software is analyzed thoroughly in this paper. Our experimental results show that AES-CBC achieves the higher security performance compared to AES-ECB scheme, although the speed of encryption degrades marginally. Visual appearance of test images demonstrates the superior confusion and diffusion properties of AES-CBC since there is full scrambling.

As future work, we are investigating further memory optimization using pre-computed tables to enhance the round operations, such as *SubByte/InvSubByte,* by exploiting similarities between encryption and decryption. As the AES encryption scheme becomes more widely used, the concept of mixed software and hardware design is also a growing new area of interest.

## 7. References

[1] IEEE Standard for Information Technology – Telecommunications and Information Exchange between Systems – Local and Metropolitan Networks – Specific requirements, Part II Amendment 6: Medium Access Control (MAC) Security Enhancements, IEEE Std 802.11i – 2004.

[2] R. Struik, "Formal specification of the CCM* Mode of operation." Doc no. IEEE 15-04-0537-00-004b.

[3] Rainer Leupers, "Code Optimization Techniques for Embedded Processors –Methods, Algorithms, and Tools", Kluwer Academic Publishers, 2000.

[4] Duncan S. Wong, Hector Ho Fuentes, Agnes H. Chan, "The Performance Measurement of Cryptographic Primitives on Palm Devices" In the Proceedings of the $17^{th}$ Annual Computer Security Applications Conference, December 2001.

[5] NIST-FIPS Standard "Announcing the Advanced Encryption Standard (AES)," in Federal Information Processing Standards Publication no. 197, November 2001.

[6] J. Daemen, V. Rijmen, "The block cipher Rijndael", Proceedings of the Third International Conference on smart card Research and Applications, CARDIS'98, Lecture Notes in computer Science, vol.1820, Springer, Berlin, 2000, pp.277-284.

[7] G. Bertoni, L. Breveglieri, P. Fragneto, M. Macchetti and S. Marchesin, " Efficient Software Implementation of AES on 32-bit platforms" in Cryptographic Hardware and Embedded Systems CHES 2002, pp. 159-171.

[8] K. Aoki1 and H. Lipmaa, "Fast implementations of AES candidates", Third AES Candidate Conference (New York City, USA), April 2000.

[9] R. Doomun, KMS Soyjaudah, D. Bundhoo, "Energy Consumption and Computational Analysis of Rijndael-AES", Third IEEE International Conference in Central Asia on Internet The Next Generation of Mobile, Wireless and Optical Communications Networks, ICI 2007, September 26-28, 2007.

[10] R. Doomun, KMS Soyjaudah, "Analytical Comparison of Cryptographic techniques for Resource Constrained Wireless Security". http://ijns.nchu.edu.tw/ International Journal of Network Security. (accepted Feb 2008)

[11] J. Daemen., Rijmen, V., "AES Proposal: Rijndael", AES submission, 1998. Available at <http://csrc.nist.gov/encryotion/aes/aes_home.htm>.

[12] William Stallings, "Cryptography and Network Security", 2$^{nd}$ Edition, Prentice Hall, 1999.